\documentclass[a4paper,twoside]{article}

\usepackage{epsfig}
\usepackage{subfigure}
\usepackage{calc}
\usepackage{amssymb}
\usepackage{amstext}
\usepackage{amsmath}
\usepackage{amsthm}
\usepackage{multicol}
\usepackage{pslatex}
\usepackage{apalike}
\usepackage[hyphens]{url}
\usepackage{ctable} 
\usepackage{xspace} 
\usepackage{SCITEPRESS}     

\newcommand{\setype}[1]{{\tt #1}\xspace}
\newcommand{\p}{\phantom{0}}

\newif\ifanonymous
\newif\ifnotanonymous
\newif\ifabridged
\newif\ifnotabridged

\ifdefined\isabridged
\abridgedtrue
\fi

\ifabridged\notabridgedfalse
\else\notabridgedtrue
\fi

\ifdefined\isanonymous
\anonymoustrue
\fi

\ifanonymous\notanonymousfalse
\else\notanonymoustrue
\fi

\begin{document}

\title{Characterizing SEAndroid Policies in the Wild}

\ifnotanonymous
\author{\authorname{Elena Reshetova\sup{1}, Filippo Bonazzi\sup{2}, Thomas Nyman\sup{2}, Ravishankar Borgaonkar\sup{2} and N. Asokan\sup{3}}
\affiliation{\sup{1}Intel OTC and Aalto University, Finland}
\affiliation{\sup{2}Aalto University, Finland}
\affiliation{\sup{3}Aalto University and University of Helsinki, Finland}
\email{\sup{1}elena.reshetova@intel.com, \sup{2}\{name.surname\}@aalto.fi, \sup{3}asokan@acm.org}
}
\fi


\abstract{Starting from the 5.0 Lollipop release all Android processes must be run inside confined SEAndroid access control domains.
As a result, Android device manufacturers were compelled to develop SEAndroid expertise in order to create policies for their device-specific components.
In this paper we analyse SEAndroid policies from a number of 5.0 Lollipop devices on the market, and identify patterns of common problems we found.
We also suggest some practical tools that can improve policy design and analysis.
We implemented the first of such tools, SEAL.}

\onecolumn \maketitle \normalsize \vfill

\section{\uppercase{Introduction}}
\label{sec:introduction}
\noindent During the past decade Android has gained a considerable share of the mobile device market.
However, at the same time the number of malware and various exploits available for Android has also been increasing~\cite{zhou2012dissecting,smalley2013security}.
Many classical Android exploits, such as GingerBreak and Exploid~\cite{smalley2013security}, attempt to target system daemons that run with elevated, often unlimited, privileges.
Once such a daemon is compromised, the whole Android OS usually becomes compromised and the attacker is able to get permanent root privileges on the device.
Since the Android permission system, which relies on Linux Discretionary Access Control (DAC), cannot protect from such attacks, a new Mandatory Access Control (MAC) mechanism has been introduced.
SEAndroid~\cite{smalley2013security} is an Android port of the well-established SELinux MAC mechanism~\cite{smalley2001implementing} with some Android-specific additions and modifications.
In SELinux, security decisions are taken according to a policy: the reference policy for SEAndroid was created from scratch and is maintained as part of the Android Open Source Project (AOSP)~\footnote{\url{source.android.com}}.

Starting from the Android 5.0 Lollipop release, the Android compliance requirements have mandated that every process must be run inside a confined SEAndroid domain with a proper set of access control rules defined.
This has put many Android Original Equipment Manufacturers (OEMs) in the difficult position of enabling SEAndroid in enforcing mode on their devices with a set of fully configured access control domains.
While the reference SEAndroid policy is provided by AOSP, any OEM customization to the reference AOSP device design results in a need for SEAndroid policy modifications.
Writing well-designed SELinux policies requires expertise; this difficulty, together with high time-to-market pressure, can possibly lead to the introduction of mistakes and outright vulnerabilities in modified SEAndroid policies deployed in OEM Android devices.

In this paper, we conduct a systematic manual analysis of several available SEAndroid 5.0 Lollipop OEM policies and identify common patterns and mistakes.
We find that OEM modifications can render policies less strict, resulting in a wider attack surface for potential vulnerabilities.
Based on these findings, we identify a number of practical tools that can assist SEAndroid policy designers and researchers to analyze and improve SEAndroid policies.
We also provide an initial implementation of one such tool, \textit{SEAL}.
To the best of our knowledge, this is the \emph{first comparative study} of SEAndroid policies from real-world devices.

\section{\uppercase{Background}}
\noindent
\subsection{SELinux}
SELinux~\cite{smalley2001implementing} is a well-established MAC mechanism available for Linux-based distributions.
It was the first MAC for mainstream Linux with its initial release in 1998.
SELinux has been implemented in the Linux kernel following the Flask architecture~\cite{spencer1999flask}, where the policy enforcement code, known as Linux Security Module (LSM) Framework, and the policy decision-making code are separated: this allows other MAC modules, such as AppArmor~\cite{bauer2006paranoid} or Smack~\cite{schaufler2008smack}, to utilize the same policy enforcement code.
\ifnotabridged
Conceptually, the LSM framework is a set of hooks in security-sensitive places of the Linux kernel, allowing a certain operation, like opening a file, to succeed only if the hook function returns a positive answer.
Compared to other MAC modules in the upstream Linux kernel, SELinux can be considered to be the most fine-grained.
It is also widely considered to be the most difficult to understand and manage, because it does not have any learning mode as AppArmor or a simple minimal policy like Smack~\cite{selinuxcomparison}.
Despite this, a number of distributions such as Red Hat Enterprise Linux (RHEL) and Fedora have SELinux enabled and in enforcing mode by default, with pre-defined security policies.

SELinux actually supports a number of independent MAC mechanisms on its own, such as Domain/Type Enforcement~\cite{badger1995practical}, Role-Based Access Control~\cite{sandhu1996role} and Multilevel Security (MLS)~\cite{qiu1985trusted}.
In principle, if one were to write an SELinux policy from scratch, it is possible to use only one or a combination of these mechanisms.
However the reference SELinux policy utilizes all three of them, making it harder for a novice to become proficient in understanding and managing it.
The SELinux Domain/Type Enforcement mechanism assigns a \setype{type} to each subject or object in the system; a subject's type is also known as \setype{domain}.
\else

The main MAC mechanism in SELinux is Domain/Type Enforcement~\cite{badger1995practical}, which assigns a \setype{type} to each subject or object in the system; a subject's type is also known as \setype{domain}.
\fi
A subject running in \setype{domain} can only access an object belonging to \setype{type} if there is an \setype{allow} rule in the policy of the following form:

\begin{center}
\setype{allow domain type : class permissions}
\end{center}

where \setype{class} represents the nature of an object such as file, socket or property, and \setype{permissions} represent the types of operation on this object that are being controlled, like \setype{open}, \setype{write}, \setype{set} etc.
Subjects can change their \setype{domain} if a corresponding type transition rule is defined.
For example, if a process executes a new binary, it is possible for the resulting process to run in a different \setype{domain}. Such a transition rule will be represented as:

\begin{center}
\setype{type\_transition olddomain type:process newdomain}
\end{center}

where \setype{type} denotes the type of a binary that should be executed in order for the transition to happen.
We will use the term \textit{process transition} in the future to refer to such a rule.
In practice there are a number of additional rules that are needed in order to make the transition happen, but we leave them out here for the sake of simplicity.
Another type of transition can occur if an object is created and its \setype{type} should differ from the \setype{type} of the object's parent.
For file creation inside an existing directory such a rule can be represented as:

\begin{center}
\setype{type\_transition domain oldtype:dir newtype}
\end{center}

where \setype{domain} denotes the domain of the subject creating the file, \setype{oldtype} represent the \setype{type} of the directory where the file is being created and \setype{newtype} denotes the \setype{type} that the new file should be assigned.

In addition, the SELinux policy language has the following notions:

\begin{itemize}
	\item An \setype{attribute} is a way to refer to sets of types and domains. It is used to express type hierarchies and rule inheritance. For example, SELinux policies on Android define an \setype{app} attribute consisting of common rules for all platform applications such as the ability access the device display.
	\item \textit{Initial \setype{Security Identifier (SID)}} are types that should be assigned by default to subjects and objects during system initialization, such as for example \setype{kernel} and \setype{init}.
	\item \setype{genfs} contexts define types that should be assigned to objects residing in special filesystems, such as \setype{proc}, \setype{debugfs} and \setype{ecryptfs}.
\end{itemize}

\subsection{SEAndroid}
The SELinux port to Android, SEAndroid~\cite{smalley2013security}, was mostly based on SELinux code with some additional LSM hooks to support Android-specific mechanisms, such as Binder Inter Process Communication (IPC).
However, the SEAndroid reference policy was written from scratch due to 1) a desire for a simpler and a smaller policy and 2) the big difference between the userspace layers of Android and a standard Linux distribution.

\ifnotabridged
Currently, the SEAndroid policy only uses Domain/Type Enforcement from SELinux.
\fi
SEAndroid classes and permissions are mostly the same as on SELinux, with some Android-specific additions like the \setype{property} class for the Android init-based property service and the \setype{keystore\_key} class for the Android keystore key object.

Native services and daemons are assigned SEAndroid domains based on filesystem labeling or direct domain declaration in the service definition in the \setype{init.rc} file.
In turn, applications are assigned domains based on the signature of the Android application package file (\setype{.apk}).
There are a number of predefined application domains, like \setype{system\_app}, \setype{platform\_app} and \setype{untrusted\_app}.
OEMs are able to create additional domains if needed.

One new notion that SEAndroid has is the presence of \setype{neverallow} rules in the source policy.
A \setype{neverallow} rule specifies that certain accesses should never be allowed by the policy.
For example, the following \setype{neverallow} rule asserts that only processes running in the \setype{init} domain should be able to modify security-sensitive files in the \setype{proc} filesystem:

\begin{center}
\setype{neverallow \{domain -init\} proc\_security:file \{append write\}}
\end{center}

If one tries to add a rule that conflicts with a \setype{neverallow} rule, the policy compilation fails.
\ifnotabridged
Since the Android 5.1 Lollipop release, OEMs are not allowed to modify the AOSP set of \setype{neverallow}rules, which makes it a very strong enforcement point in guiding OEM policy modifications.
\fi

SEAndroid was initially added to the AOSP codebase for Android 4.3 back in 2012; at that point, it was configured in permissive mode.
In Android 4.4, SEAndroid was switched to enforcing mode: however, most domains were left in permissive mode, apart from a number of core AOSP domains such as \setype{init} and \setype{vold}.
Android 5.0 Lollipop eventually required every single process to be put in an enforcing domain, effectively extending the enforcement to the whole system.

\subsection{OEM Modifications to the AOSP SEAndroid Reference Policy}
Default AOSP services, processes and applications are already covered by the AOSP SEAndroid reference policy.
Typically, OEM Android devices are highly customized with their own specific drivers, new services, processes and filesystem mounts.
In order for these custom components to work, appropriate additions must be made to the SEAndroid reference policy.
OEMs were allowed to make additions to the SEAndroid reference policy right from the start, but very few of them actually did in Android 4.3 and 4.4: and in fact, the resulting policy was stricter than the AOSP reference policy.
The stringent requirements of Android 5.0 Lollipop, however, forced all OEMs to deploy comprehensive policies defining complete rules for their own custom services: this turned out to be a challenging task for most.
The inherent difficulty of incorporating SEAndroid in the development process, combined with high time-to-market pressure, has resulted in the introduction of anti-patterns, mistakes and potential vulnerabilities in OEM policies.

\section{\uppercase{SEAndroid Policies in Deployed OEM Devices}}
\noindent
\subsection{Statistical Analysis}
\label{sec:stat}
We collected 8 policy files from non-rooted, off-the-shelf commercial Android 5.0 Lollipop devices by different manufacturers.
Table~\ref{tab:comp} shows the comparison of all basic policy attributes and characteristics with regards to the Android 5.0 Lollipop AOSP SEAndroid reference policy \ifnotabridged and the Fedora 22 minimal desktop policy \fi in the following categories:

\begin{itemize}

\item \textbf{Policy size.} All OEMs increased the policy size, by factors ranging from 1.1 up to 3.2.
\ifnotabridged
The reference Fedora policy is still bigger than the biggest OEM policy by a factor of 10, which is due to a number of reasons.
First, Fedora's SELinux policy uses Role-Based Access Control with 14 roles, 8 users and 414 role transitions, as well as MLS categories with 5466 MLS range transitions.
Second, policy in Fedora actively uses SELinux policy booleans (293 policy booleans are defined) in order to be able to dynamically adjust policy behavior, as well as \setype{portcons} statements that allow to assign security contexts to UDP or TCP ports.
In addition, Android has fewer components and processes running in userspace compared to a typical desktop system such as Fedora.
\fi

	\item \textbf{Types, domains, type transitions and domain transitions.}
The overall ratio of newly added domains to newly added types ranges between 4.7 and 6.5, which is very close to the ratio in AOSP itself (6.3).
We conjecture that OEMs tend to add slightly simpler domains, with fewer types per each domain defined.
The ratio of newly added process transitions to newly added domains is close to 1; this indicates that OEMs add simple domains, with only one process transition to these domains either from the \setype{init} domain (upon system startup) or from the parent process domain (usually upon execution of processes such as \setype{shell} or \setype{toolbox}).
This ratio is 1.4 for LG G3, due to a number of newly defined domains and having two or more transitions to the \setype{shell}, \setype{toolbox}, \setype{dumpstate} and \setype{logcat} domains.
New type transitions are mostly used by OEMs for \setype{tmpfs} types, as in the following example:

\begin{center}
\setype{type\_transition aal tmpfs : file aal\_tmpfs}
\end{center}

	\item \textbf{Allow rules.}
	The ratio of total number of allow rules to total number of types varies between 10.9 and 13.1, with the exception of 14.8 for Motorola G and 18.7 for LG G3; the ratio for AOSP is 12.0.
The numbers for Motorola and LG are comparatively excessive, and may indicate overly permissive policies; this may be due to the use of tools to automatically generate policies from system logs.

	\item \textbf{Attributes.}
	Only Samsung and Sony define new attributes.
Sony adds only one, probably related to the system update process.
In contrast, Samsung adds many new attributes, which seem to be auto-generated and most probably used for policy optimization.
The rest of the OEMs add separate domains for their services and applications, and do not introduce any new domain hierarchies: this may imply unfamiliarity with the use of policy hierarchies.

	\item \textbf{Classes, permissions and initial SIDs.}
	OEMs do not modify the default set of SEAndroid \setype{classes} (86), \setype{permissions} (267) or initial \setype{SIDs} (27): this is to be expected, since they represent interfaces and objects recognized and supported by SEAndroid.
The only change we observed was for Samsung S6, that had 4 more permissions defined: \setype{delete\_as\_user}, \setype{get\_by\_uid}, \setype{insert\_as\_user} and \setype{set\_max\_retry\_count}, all granted on the \setype{keystore\_key} class.
The most probable reason for such additions is Samsung's implementation of the keystore and its API, which requires specific permissions.
\ifnotabridged
Compared to Fedora, SEAndroid has three more classes due to the Binder IPC support.
\fi

	\item \textbf{\setype{genfs} contexts.}
	The primary reason for the addition of \setype{genfs} contexts is that most OEMs have additional mount points and filesystems on their devices, which by default would be labeled as \setype{unlabeled} unless a proper \setype{genfs} context for it is specified.
An AOSP \setype{neverallow} rule prohibits any OEM domain from creating files with this type: this restriction has forced OEMs to define proper types for their new mount points.

\end{itemize}

\begin{table*}[t]
 \centering
	\caption{Policy comparison and complexity.}
	\label{tab:comp}
	\resizebox{\textwidth}{!}{%
	\begin{tabular}{ | c | c | c | c | c | c | c | c | c | c | }

   \hline
    &  \textbf{\textit{size (KB)}} & \textbf{\textit{types}}  & \textbf{\textit{domains}} & \textbf{\textit{type trans}} & \textbf{\textit{process trans}}  &  \textbf{\textit{allow rules}} & \textbf{\textit{attributes}} & \textbf{\textit{genfs contexts}} & \textbf{\textit{untrusted\_app rules}}  \\  \hline

    \ifnotabridged \textit{Fedora}&3851&4638& 802      & 16917    & 11795     & 102645         & 357	    & 101    & n/a    \\ \hline \fi

   \textit{AOSP}       & 117 & 341         & 54        & 95       & 41        & 4096	       & 21         & 30     & 33     \\ \specialrule{.1em}{.05em}{.05em} \specialrule{.1em}{.05em}{.05em}

   \textit{LG Nexus 5} & 134 &\p416, \p+75 &\p65, \p+11&158, \p+63&\p51, \p+10&\p4972, \p\p+876&21\p\p\p\p  &32, \p+2& 33\p\p\p\p\\ \hline

   \textit{Intel}      & 127 &\p393, \p+52 &\p65, \p+11&115, \p+20&\p51, \p+10&\p4748, \p\p+652&21\p\p\p\p  &32, \p+2& \p38, \p\p+5 \\ \hline

   \textit{HTC M7}     & 181 &\p621, +280  & 106, \p+52& 213, +118&\p95, \p+54&\p7587, \p+3491 &21\p\p\p\p  &34, \p+4& \p46, \p+13\\ \hline

   \textit{Motorola G} & 193 &\p590, +249  &\p92, \p+38& 199, +104&\p83, \p+42&\p8753, \p+4657 &21\p\p\p\p  &33, \p+3& 33\p\p\p\p\\\hline

   \textit{LG G3}      & 302 &\p851, +510  & 149, \p+95& 340, +245&180, +139  &15921, +11825   &21\p\p\p\p  &45, +15 &168, +135\\ \hline

   \textit{Intex Aqua} & 230 &\p900, +559  & 142, \p+88& 266, +171&128, \p+87 &\p9824, \p+5728 &21\p\p\p\p  &37, \p+7& \p44, \p+11\\ \hline

   \textit{Samsung S6} & 370 &1102, +761   & 215, +161 & 430, +335&180, +139  &14412, +10316   &158, +137   &43, +13 & \p81, \p+48\\ \hline

   \textit{Sony Xperia}& 218 &\p793, +452  & 139, \p+85& 265, +170&113, \p+72 &\p9308, \p+5212 &\p22, \p\p+1&37, \p+7& \p42, \p\p+9 \\ \hline

   \end{tabular}}

\begin{flushleft}
{\footnotesize + \textit{denotes the number of additions compared to AOSP}}
\end{flushleft}
\end{table*}

\ifnotabridged
\begin{figure*}[htbp]
  \centering
	{\includegraphics[width=0.8\textwidth]{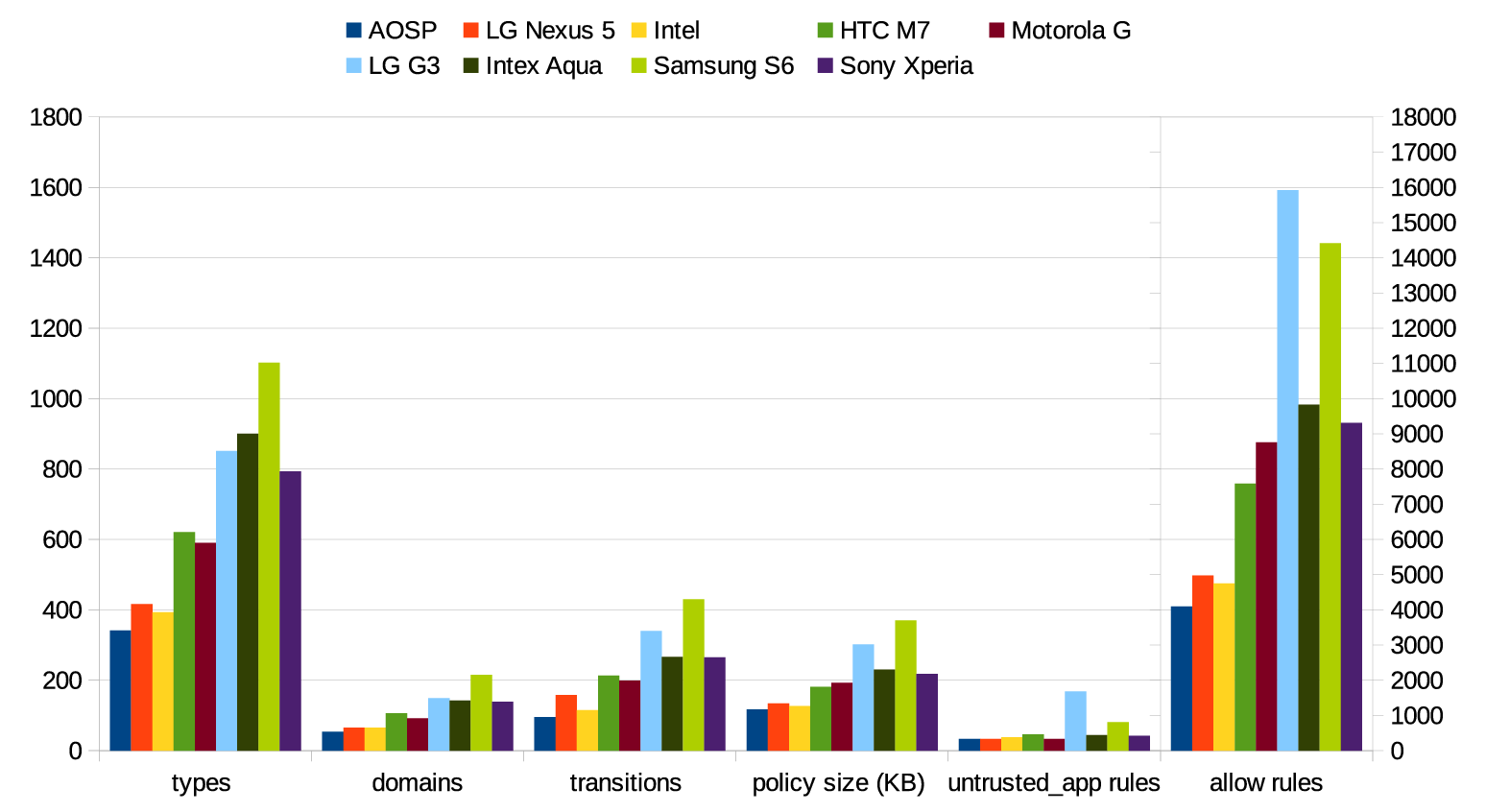}}
  \caption{Policy comparison and complexity.}
  \label{fig:comp}
\end{figure*}
\fi

\subsection{Systematic Manual Analysis}
\label{sec:sma}
We manually searched each policy for OEM misconfigurations: we used existing tools for SELinux policy analysis, which we found to be cumbersome.
Our primary tool was \setype{apol}~\cite{setools3}, a GUI tool that allows the user to load a binary policy and examine it by specifying various filters.
However, \setype{apol} was not suitable for comparing two policies: it was necessary to run two instances of \setype{apol} simultaneously, manually insert the same queries into both and examine the differences between the outputs.
Another tool was \setype{sediff}~\cite{setools3}, which can do basic policy comparison but does not allow filtering based on specific types or domains.

To make our manual analysis tractable, we identified three sets of types that we consider important to check.
The first set comprises core Android and security-sensitive domains, such as \setype{init}, \setype{vold}, \setype{keystore}, \setype{tee}, as well as types that protect access to security-sensitive areas of the filesystem, such as \setype{proc\_security}, \setype{kmem\_device} and \setype{security\_file}.
The second is the set of default types that would be assigned to an object upon its creation unless a concrete type is specified in one of the policy files.
The third is the set of types that would be assigned to untrusted code and its data, primarily the \setype{untrusted\_app} domain.

Analyzing these sets of types and the associated rules, we discovered the following patterns across many devices from different OEMs.

\subsubsection{Overuse of Default Types}
\label{sec:udt}
As mentioned above, a default type is one that is assigned to an object upon creation unless a dedicated type for it is specified in the policy files: examples include \setype{unlabeled}, \setype{device}, \setype{socket\_device}, \setype{default\_prop} and \setype{system\_data\_file}.
Table~\ref{tab:default} shows that in many cases OEMs overuse the default SEAndroid object types.
For example, compared to AOSP, HTC M7 has 10 new rules allowing various system daemons, such as \setype{healthd}, \setype{netd}, \setype{vold}, \setype{mediaserver}, \setype{wpa}, \setype{system\_server}, to set system properties with the default type \setype{default\_prop}.
In practice, this means that some of the system properties belonging to these components end up labeled as \setype{default\_prop}.
Similarly, HTC M7 has 13 more rules granting various system daemons (\setype{rild}, \setype{mediaserver}, \setype{thermal-engine}, \setype{sensors}, \setype{thermald}, \setype{system-server}, \dots) \setype{write} access to the default \setype{socket\_device} object type.
The only exceptions when OEM actually reduced the number of rules with regards to default types are LG Nexus 5 and Motorola G policies, where a rule for \setype{logd} to access /setype{device} was removed, and Samsung S6 policy where a set of rules for unused \setype{unconfined} domain was removed.
Below are concrete example rules from different OEMs to show the usage of default types:

\begin{center}
\setype{allow thermald socket\_device : sock\_file \{write create setattr unlink\}}
\end{center}
\begin{center}
\setype{allow mediaserver default\_prop : property\_service set}
\end{center}
\begin{center}
\setype{allow untrusted\_app unlabeled : dir \{ioctl read getattr search open\}}
\end{center}
\begin{center}
\setype{allow untrusted\_app unlabeled : filesystem getattr}
\end{center}

\begin{table*}[t]
\small
 \centering
	\caption{Usage of default types by OEMs.}
	\label{tab:default}
   \begin{tabular}{| c | c | c | c | c | c | }

   \hline
    & \textbf{\textit{unlabeled}} & \textbf{\textit{socket\_device}} & \textbf{\textit{device}} & \textbf{\textit{default\_prop}} & \textbf{\textit{system\_data\_file}} \\ \hline

   \textit{AOSP}       &25        &4         & 18         & 0          & 42 \\ \specialrule{.1em}{.05em}{.05em} \specialrule{.1em}{.05em}{.05em}

   \textit{LG Nexus 5} &25\p\p\p\p&\p7, \p+3 &17, \p--1 & \p0\p\p\p\p& \p45, \p\p+3 \\ \hline

   \textit{Intel}      &27, \p+2  &\p5, \p+1 &24, \p+6  & \p3, \p+3  & \p54, \p+12 \\ \hline

   \textit{HTC M7}     &31, \p+6  &17, +13   &26, \p+8  & 10, +10    & \p68, \p+26 \\ \hline

   \textit{Motorola G} &31, \p+6  &\p7, \p+3 &17, \p--1 & \p2, \p+2  & \p62, \p+20 \\\hline

   \textit{LG G3}      &42, +17   &21, +17   &28, +10   & \p3, \p+3  & 120, \p+78\\ \hline

   \textit{Intex Aqua} &33, \p+8  &\p8, +14  &23, \p+5  & \p0\p\p\p\p& 108, \p+66\\ \hline

   \textit{Samsung S6} &24, \p--1   &22, +18   &46, +28   & \p1, \p+1  & 204, +162\\ \hline

   \textit{Sony Xperia}&25\p\p\p\p&15, +11   &21, \p+3  & \p1, \p+1  & \p57, \p+15 \\ \hline

   \end{tabular}

\begin{flushleft}
{\footnotesize $\pm$ \textit{denotes the number of additions/removals compared to AOSP}}
\end{flushleft}
\end{table*}

\ifnotabridged
\begin{figure*}[htbp]
  \centering
	\includegraphics[width=0.8\textwidth]{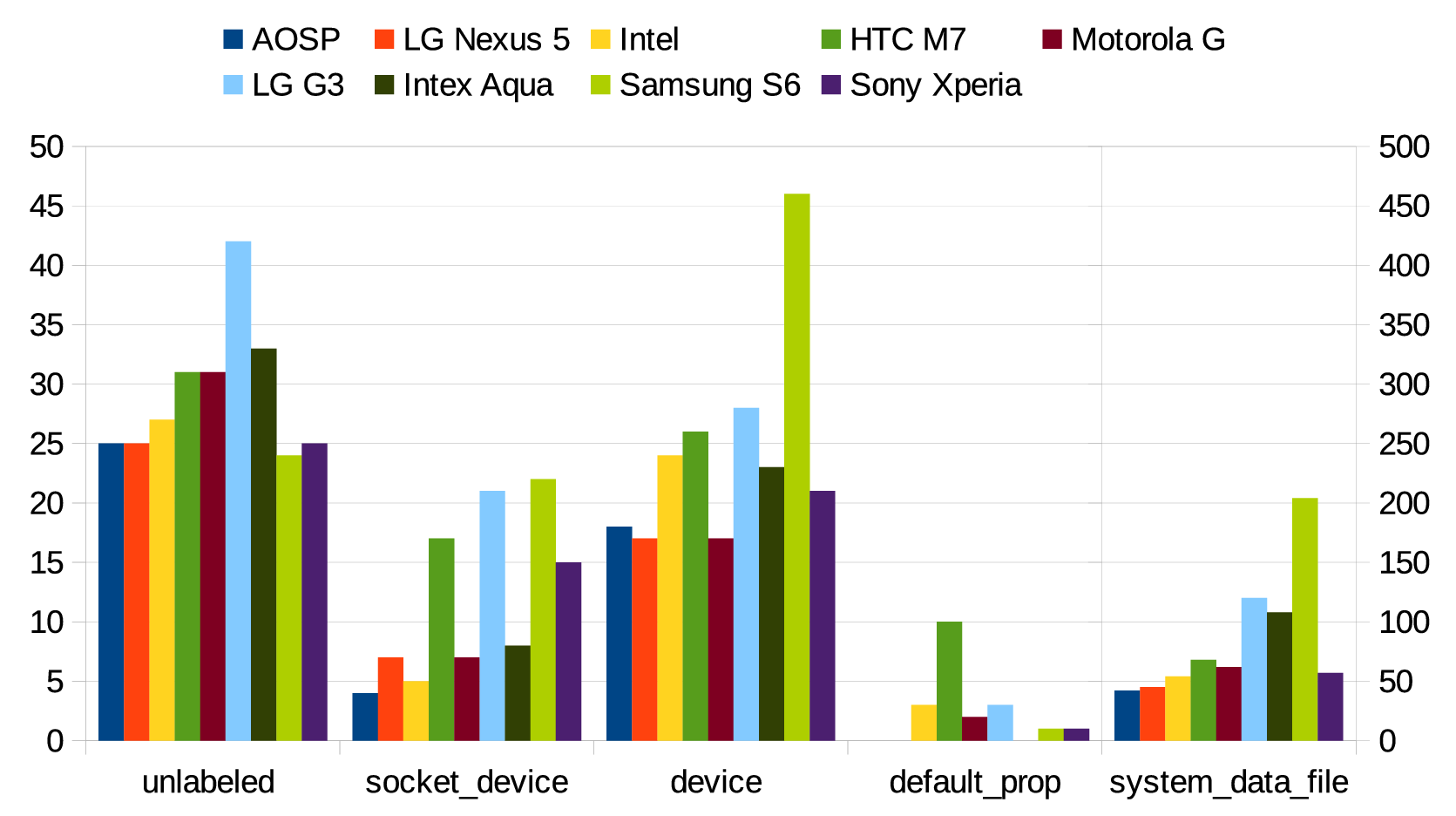}
  \caption{Usage of default types by OEMs.}
  \label{fig:default}
\end{figure*}
\fi

Plausible reasons for OEMs to use default types include the fact that objects are automatically assigned default types, and the common practice of using tools like \setype{audit2allow}~\cite{selinuxuserspace} which parse audit logs and automatically create new allow rules to permit denied accesses.

There are two main consequences of such mistakes.
Foregoing distinct, dedicated types in favor of default types means that different, unrelated resources are collected under a common label: domains with access to said label thus get wider access rights than actually needed.
This is undesirable, as it violates the principle of least privilege.
The second, more severe, consequence is that some untrusted domains might be given access to default labeled sensitive objects.

Fortunately, we did not find examples of such cases in the policies we examined, apart from the example above where \setype{untrusted\_app} is given some access to \setype{unlabeled} filesystem objects; however, the possibility of such mistakes remains.

Google is actively trying to address this problem by fine-tuning the set of \setype{neverallow} rules in the AOSP reference policy.
Starting from Android 5.1, OEMs are not allowed to modify this set. For example, it is not possible anymore for an OEM domain to set default properties or access block devices.

\subsubsection{Overuse of Predefined Domains}
Another observed trend is that typically OEMs do not define separate domains for specific system applications, but tend to place them either in \setype{system\_app} or in \setype{platform\_app} domains.
Consequently, these domains accumulate a lot of \setype{allow} rules that are shared by all system or platform applications.
Moreover, if many applications are pre-installed in the same domain, SEAndroid cannot prevent privilege escalation attacks or unauthorized data sharing by such apps~\cite{smalley2013security}.
As an example, let us consider the pre-installed McAfee anti-virus application on LG devices.
It runs in the \setype{system\_app} domain, which contains more than 900 associated allow rules in the LG G3 policy compared to 46 in the AOSP one.
It is quite difficult to identify specific rules that were added to the \setype{system\_app} domain because of the McAfee application, given that many other applications run in this domain.
However, by analyzing the permissions of the same McAfee application in Google Play Store, we observed corresponding SEAndroid policy rules in the \setype{system\_app} domain, such as access to the telephony functionality, camera and several types related to the filesystem, including \setype{tmpfs} types and sockets.
This might indicate that these rules were added for the McAfee application.

A solution to this problem would be to place certain powerful system applications in their own SEAndroid domains; this can be done by signing these applications with different keys and creating a mapping between these keys and target application domains.
\ifnotabridged
Application providers can then distribute their applications to OEMs signed with these keys, and get their application to run with the correct set of permissions.
\fi

\subsubsection{Forgotten or Seemingly Useless Rules}
Another common trend is the presence of rules that seem to have no effect. One example is rules of the following type present in one device:

\begin{center}
\setype{allow untrusted\_app <xyz>\_exec : file <file op>}
\end{center}

For example

\begin{center}
\setype{allow untrusted\_app tee\_exec : file \{read getattr execute open\}}
\end{center}

Since no corresponding process transition rule from the \setype{untrusted\_app} domain to the \setype{tee} domain via \setype{tee\_exec} file is defined, and no \setype{execute\_no\_trans} access type is granted, a process running in the \setype{untrusted\_app} domain cannot execute a file labeled as \setype{tee\_exec}.
There are two plausible explanations.
One is the use of tools like \setype{audit2allow}~\cite{selinuxuserspace} to automatically generate rules, as discussed above.
The other is the failure to clean up rules that were tested at some point but are no longer required.
Below is an example of a vestigial rule that allows access to the debug interface of the Qualcomm KGSL GPU driver, which is itself disabled in production builds:

\begin{center}
\setype{allow untrusted\_app sysfs\_kernel\_debug\_kgsl : file \{read getattr\}}
\end{center}

\subsubsection{Potentially Dangerous Rules}
\label{sec:pdr}
When working under tight time-to-market requirements, OEMs might decide to ship less strict security policies rather than make invasive changes to their codebase.
This leads to a number of potentially dangerous rules appearing in OEM policies, like access to the \setype{procfs} security-related filesystem objects.
The rules below give \setype{read/write} permissions on such objects to a trusted \setype{hal} domain, a \setype{release\_app} domain and an \setype{untrusted\_app} domain.

\begin{center}
\setype{allow hal proc\_security : file \{write getattr open\}}
\end{center}
\begin{center}
\setype{allow release\_app proc\_security : file \{ioctl read getattr lock open\}}
\end{center}
\begin{center}
\setype{allow	untrusted\_app proc\_security : file \{read getattr open\}}
\end{center}

While processes running in the \setype{hal} system domain can be considered trusted, the first rule is undesirable because it increases the attack surface of certain interfaces (like sensitive \setype{procfs} settings) if the trusted process is compromised.
The same applies to applications put in the \setype{release\_app} domain, as in the second rule.
The third rule is even more dangerous, because it allows malicious applications running in the \setype{untrusted\_app} domain to get sensitive information, such as \setype{mmap\_min\_addr}, memory randomization parameters and kernel pointer exposure settings, that can be used for further exploits.

Another example of a potentially dangerous rule is allowing processes running in the \setype{untrusted\_app} domain to \setype{read/write} application data belonging to the \setype{system\_app} domain:

\begin{center}
\setype{allow untrusted\_app system\_app\_data\_file : file \{read write getattr\}}
\end{center}

However, since processes from \setype{untrusted\_app} and \setype{system\_app} domains will be run with different UIDS, the Linux DAC layer would guard against such arbitrary accesses unless a system application erroneously made its own files world-accessible.

In general, OEMs should have no additions to the set of rules for the \setype{untrusted\_app} domain, because any new \setype{allow} rule increases the possible attack surface for malicious \setype{untrusted\_app} applications.
However, Table~\ref{tab:comp} shows that almost all OEMs do add new rules for the \setype{untrusted\_app} domain.
\ifnotabridged
Some small additions can be explained by the fact that OEM devices are different, and, in order for a simple application to access the display or some other basic non-sensitive functionality, they might need additional rules not present in the AOSP policy.
An example is the rule below, that gives untrusted applications \setype{read/write} access to the temporary buffers of the \setype{surfaceflinger} daemon.
The actual implementation of the graphical stack on this device makes it secure to allow such access, which is needed for the applications to operate properly.

\begin{center}
\setype{allow	untrusted\_app surfaceflinger\_tmpfs:file \{read write\}}
\end{center}

However, when OEMs define tens or hundreds of new \setype{allow} rules for the \setype{untrusted\_app} domain, it is an indicator that the policy was not designed with care and may have flaws.
We found comparatively more mistakes in those OEM policies that had more \setype{allow} rules defined for the \setype{untrusted\_app} domain.
\fi

On a positive note, OEMs are aware of such mistakes; some of them have been already fixed in the subsequent Android 5.1 update.
The major reason behind these fixes was the release of the Android Compatibility Test Suite (CTS)\footnote{\url{source.android.com/compatibility/cts/index.html}} version 5.1, that added tests to ensure that AOSP \setype{neverallow} rules are not violated by any process running on a device.
\ifnotabridged The initial CTS version 5.0 only checked that processes running in the AOSP domains did not violate \setype{neverallow} rules.
\fi

\subsubsection{Discussion}
We found several problematic patterns in the Android 5.0 OEM SEAndroid policies we examined.
We conjecture that the reason for their presence is the relative unfamiliarity with SEAndroid.
Google utilizes the set of \setype{neverallow} rules in order to try to prevent OEMs from making security mistakes.
However, while this approach might prevent some mistakes, it can also create difficulties for OEMs.
For example any Global Platform-enabled TEE design~\footnote{\url{globalplatform.org}} will likely end up with \setype{untrusted\_app} applications needing to access a kernel driver for their memory referencing.
In Android 5.1 this conflicts with the existing \setype{neverallow} rules, and as a result OEMs are forced to come up with a workaround.
In the next section we propose a set of tools that can further help OEMs to avoid security mistakes and at the same time do not imply any restrictions on OEMs.

\section{\uppercase{New tools for SEAndroid }}
Although our systematic manual analysis unearthed some problem areas in the policies we analyzed, the process was cumbersome and time-consuming using the currently available tools.
Based on our experience, we argue that new tools or new functionality in existing tools are necessary to aid both OEMs and security researchers to create and analyze SEAndroid policies effectively.
We identify several such desirable tools below. We have implemented the first on the list (live policy analyzer) and are working on the rest.

\subsection{Live Policy Analyzer}
\begin{figure}[ht]
  \centering
    \includegraphics[width=0.4\textwidth]{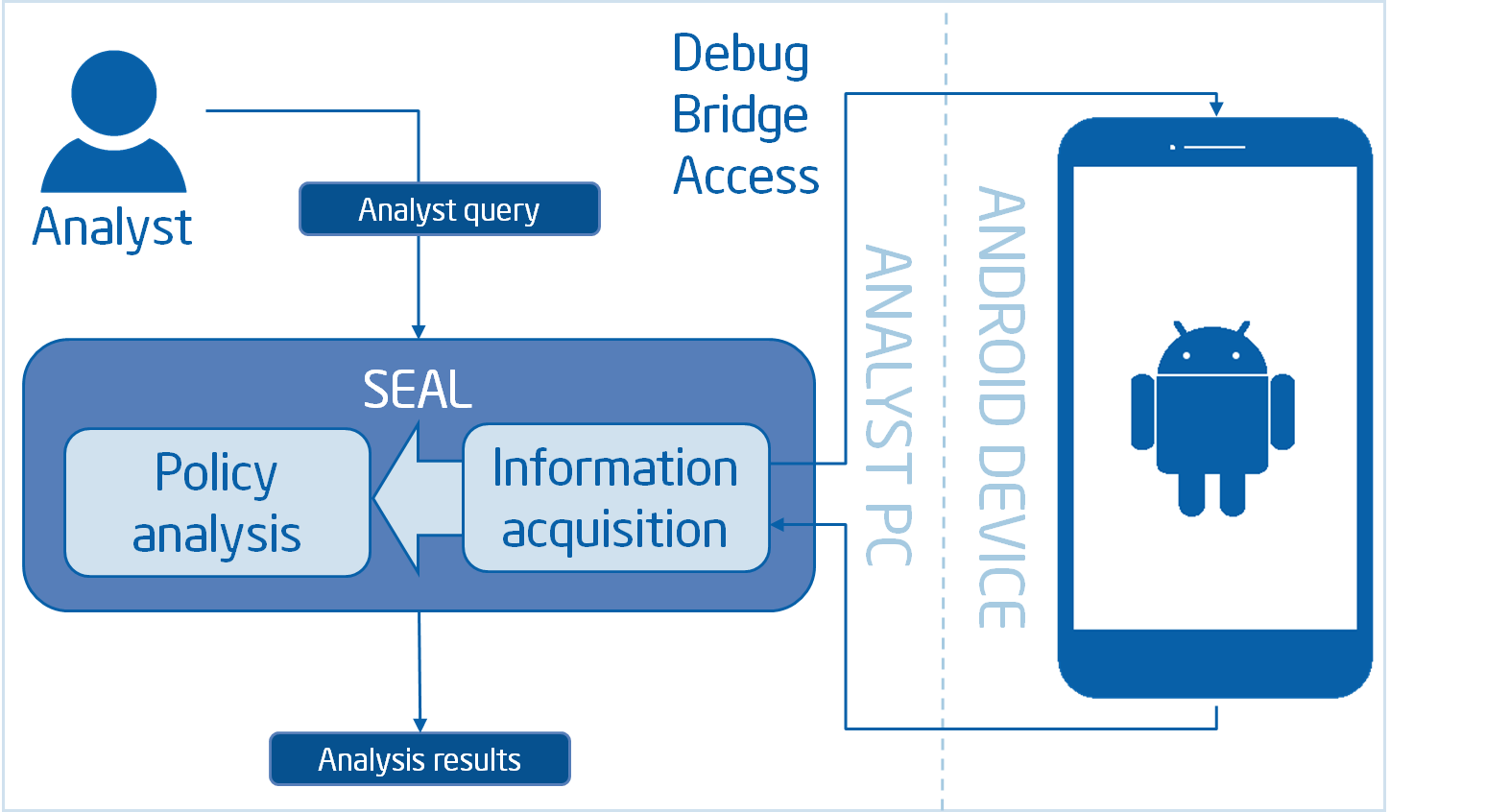}
  \caption{SEAL tool architecture}
  \label{fig:seal}
\end{figure}

Existing SELinux policy analysis tools focus solely on the policy itself, and do not address the question of how the policy rules apply to a specific target device.
A tool that can answer questions like ``what files can a specified process on a device access?'' or ``what processes on a device can access a specified file?'' would be very useful for the analyst.
We developed SEAndroid Live device analysis tool (SEAL)\footnote{\url{github.com/seandroid-analytics/seal}} for this purpose.
SEAL allows different queries that take into account not only the SEAndroid policy loaded on the device, but also the actual device state, i.e. running processes and filesystem objects.
SEAL offers command line and GUI interfaces, and queries the device over the \setype{adb} interface.
In order to obtain results about the entire device filesystem, the target device has to be either rooted or running an engineering build.
Figure~\ref{fig:seal} shows the architecture of SEAL.

\subsection{Policy Decompilation Tool}
One of the main problems during our manual analysis was the lack of a tool to easily identify and analyze changes that an OEM made to the default AOSP policy.
Current tools like \texttt{apol} and \texttt{sediff} are not directly suited for this task, as described in Section~\ref{sec:sma}.
It would be very beneficial to have a tool able to transform a binary policy into a set of source files organized similarly to the AOSP SEAndroid source policy.
In this case, it would not only be possible to perform manual analysis in a more organized and convenient manner, but also to employ standard text manipulation tools to compare or filter needed information.

\subsection{Policy Visualization Tool}
SELinux has the notion of \setype{attributes}, that allow organizing policy types and domains in a hierarchical manner.
This is a very powerful mechanism that can easily be misconfigured by mistake.
The data collected in section~\ref{sec:stat} showed that most OEMs do not create new policy attributes, perhaps due to the perceived complexity of the attribute mechanism.
A tool to visualize hierarchies induced by attributes may help an analyst better understand and make use of attributes effectively.

\subsection{Policy Analyzer}
The hardest part of our manual analysis was identifying rules that are either potentially dangerous or possibly unnecessary.
The analysis can be automated with a set of heuristic checks.
The tool can also utilize SEAL in order to make policy queries with regards to the device state.

Let us consider the following rule from section~\ref{sec:pdr} and explain how it can be automatically detected as suspicious:

\begin{center}
\setype{allow untrusted\_app tee\_exec : file \{read getattr execute open\}}
\end{center}

The tool first uses SEAL to fetch from a device all files with the \setype{tee\_exec} label.
Then it queries SEAL for all labels and DAC permissions of all higher-level directories on the path to each \setype{tee\_exec} file, and tries to determine if the \setype{untrusted\_app} domain can even reach the target file to perform the requested operations.
If the first check passes, the analyzer can further check that each requested access control type makes sense.
For example, in order for \setype{execute} to succeed given that \setype{execute\_no\_trans} access is not granted, there has to be a \setype{type\_transition} rule defined; furthermore, in order to be able to \setype{write} or \setype{read} a file, one would also normally need to have \setype{open} permission.
The policy analyzer can mark the rule as not functional if the checks fail.

In order to identify potentially dangerous rules, the policy analyzer can scan rules for possible additional usages of default types, mentioned in Section~\ref{sec:udt}, and analyze new rules associated with sensitive types, such as \setype{tee} or \setype{proc\_security}, or untrusted domains, such as the \setype{untrusted\_app} domain.

One use of the policy analyzer is its integration into OEMs' automatic build systems, in order to consistently verify that the policy does not contain unreachable or potentially dangerous rules, and that it is optimized with regards to the usage of attributes and types.
This would provide value for OEMs, since policy additions might be made by different development teams, possibly without detailed knowledge of SEAndroid.
The output of the tool can be further analyzed manually by a person with detailed knowledge of SEAndroid, in order to reject or accept suggested modifications.

\section{\uppercase{Related Work}}
\label{sec:related-work}
Several tools and methods originally designed for SELinux are relevant to the new mobile environment.

The \textit{de facto} standard for handling SELinux policies in text and binary format is the SETools library~\cite{setools3}: this contains the aforementioned \texttt{apol} and \texttt{sediff} tools, which can be used interchangeably on SELinux and SEAndroid.

Formal methods have been used for SELinux policy analysis.
Gokyo~\cite{jaeger2003analyzing} is a policy analysis tool designed to identify and resolve conflicting policy specifications.
Usage of the HRU security model~\cite{Harrison:1976:POS:360303.360333} has been proposed as an approach to SELinux policy analysis~\cite{amthor2011model}.
Information flow analysis has been applied to SELinux policies~\cite{guttman2005verifying}.
These analysis methods are not SELinux-specific, and can be easily adapted to SEAndroid.

Some researchers have applied information visualization techniques to SELinux policy analysis~\cite{clemente2012sptrack}, also in combination with clustering~\cite{marouf2011segrapher}.
These techniques are also system-agnostic, and we may use them in future SEAndroid tools.

SELinux policy generation and refining tools are rare.
Polgen, a tool for semi-automated SELinux policy generation based on system call tracing~\cite{sniffen2006guided}, appears to be no longer in active development.
The SELinux userspace tools~\cite{selinuxuserspace} can generate SELinux policies.
One of these tools, \texttt{audit2allow}, is widely used to automatically generate and refine SELinux policies by converting SELinux audit messages into rules; these policies, however, are not necessarily correct, complete or secure, since the rules depend on code paths taken during execution, and there is no way to distinguish intended and possibly malicious application behavior.
These tools are used both in SELinux and SEAndroid.

There has been some research in applying Domain Specific Languages (DSL)~\cite{fowler2010domain} to SELinux policy development and verification~\cite{hurd2009policy}.
The authors proposed a tool (\texttt{shrimp}) to analyze and find errors in the SELinux Reference Policy, similar to the \texttt{Lint} tool for C.
This is similar to a tool we propose, but different in scope as it is limited to analysis of the SELinux reference policy.

The only SEAndroid-specific analysis method is based on audit log analysis with machine learning~\cite{wang15easeandroid}.
This approach is completely different from what we propose, since it relies on significant volumes of data to classify rules.

\section{\uppercase{Conclusions}}
\label{sec:conclusion}

\noindent In this paper we presented a number of common mistakes made by OEMs in their SEAndroid policies, suggesting potential reasons behind them.
As a result of this study, we identified a number of practical tools that should help OEMs and security researchers to improve SEAndroid policies.
We provided the implementation of a first tool, SEAL, and we are currently working on the rest.

\ifnotabridged
\section*{\uppercase{Acknowledgments}}
\noindent The authors would like to thank \ifnotanonymous{William Roberts} \else {****}  \fi for his valuable suggestions on the practical applicability of the proposed SEAndroid tools, and \ifnotanonymous{Jan-Erik Ekberg} \else {****} \fi for his help with policy analysis and provided TEE use cases.
\fi

\bibliographystyle{apalike}
{\small
\bibliography{sepoly}}

\begin{thebibliography}{}

\bibitem[{Ahmad, Ali}, 2014]{selinuxcomparison}
{Ahmad, Ali} (2014).
\newblock {SELinux and AppArmor: An Introductory Comparison}.
\newblock
  \url{www.scribd.com/doc/230617085/SELinux-and-AppArmor-An-Introductory-Comparison}.
\newblock Accessed: 2015-10-19.

\bibitem[Amthor et~al., 2011]{amthor2011model}
Amthor, P., Kuhnhauser, W., and Polck, A. (2011).
\newblock Model-based safety analysis of selinux security policies.
\newblock In {\em NSS}, pages 208--215. IEEE.

\bibitem[Badger et~al., 1995]{badger1995practical}
Badger, L., Sterne, D., Sherman, D., Walker, K., Haghighat, S., et~al. (1995).
\newblock Practical domain and type enforcement for {UNIX}.
\newblock In {\em Security and Privacy}, pages 66--77. IEEE.

\bibitem[Bauer, 2006]{bauer2006paranoid}
Bauer, M. (2006).
\newblock {Paranoid penguin: an introduction to Novell AppArmor}.
\newblock {\em Linux Journal}, (148):13.

\bibitem[Clemente et~al., 2012]{clemente2012sptrack}
Clemente, P., Kaba, B., Rouzaud-Cornabas, J., Alexandre, M., and Aujay, G.
  (2012).
\newblock Sptrack: Visual analysis of information flows within selinux policies
  and attack logs.
\newblock In {\em AMT}, pages 596--605. Springer.

\bibitem[Fowler, 2010]{fowler2010domain}
Fowler, M. (2010).
\newblock {\em Domain-specific languages}.
\newblock Pearson Education.

\bibitem[Guttman et~al., 2005]{guttman2005verifying}
Guttman, J.~D., Herzog, A.~L., Ramsdell, J.~D., and Skorupka, C.~W. (2005).
\newblock {Verifying information flow goals in security-enhanced Linux}.
\newblock {\em Journal of Computer Security}, 13(1):115--134.

\bibitem[Harrison et~al., 1976]{Harrison:1976:POS:360303.360333}
Harrison, M.~A., Ruzzo, W.~L., and Ullman, J.~D. (1976).
\newblock Protection in operating systems.
\newblock {\em Commun. ACM}, 19(8):461--471.

\bibitem[Hurd et~al., 2009]{hurd2009policy}
Hurd, J., Carlsson, M., Finne, S., Letner, B., Stanley, J., and White, P.
  (2009).
\newblock {Policy DSL: High-level Specifications of Information Flows for
  Security Policies}.

\bibitem[Jaeger et~al., 2003]{jaeger2003analyzing}
Jaeger, T., Sailer, R., and Zhang, X. (2003).
\newblock Analyzing integrity protection in the selinux example policy.
\newblock In {\em USENIX Security}, page~5.

\bibitem[Marouf and Shehab, 2011]{marouf2011segrapher}
Marouf, S. and Shehab, M. (2011).
\newblock {SEGrapher: Visualization-based SELinux policy analysis}.
\newblock In {\em SAFECONFIG}, pages 1--8. IEEE.

\bibitem[Qiu et~al., 1985]{qiu1985trusted}
Qiu, L., Zhang, Y., Wang, F., Kyung, M., and Mahajan, H. (1985).
\newblock Trusted computer system evaluation criteria.
\newblock In {\em National Computer Security Center}. Citeseer.

\bibitem[Sandhu et~al., 1996]{sandhu1996role}
Sandhu, R., Coyne, E., Feinstein, H., and Youman, C. (1996).
\newblock Role-based access control models.
\newblock {\em Computer}, (2):38--47.

\bibitem[Schaufler, 2008]{schaufler2008smack}
Schaufler, C. (2008).
\newblock Smack in embedded computing.
\newblock In {\em Ottawa Linux Symposium}.

\bibitem[{SELinux Project}, 2014]{selinuxuserspace}
{SELinux Project} (2014).
\newblock {Userspace tools}.
\newblock \url{github.com/SELinuxProject/selinux/wiki}.
\newblock Accessed: 2015-09-29.

\bibitem[Smalley and Craig, 2013]{smalley2013security}
Smalley, S. and Craig, R. (2013).
\newblock Security {E}nhanced ({SE}) {A}ndroid: Bringing flexible {MAC} to
  {A}ndroid.
\newblock In {\em NDSS}, volume 310, pages 20--38.

\bibitem[Smalley et~al., 2001]{smalley2001implementing}
Smalley, S., Vance, C., and Salamon, W. (2001).
\newblock {Implementing SELinux as a Linux security module}.
\newblock {\em NAI Labs Report}, 1(43):139.

\bibitem[Sniffen et~al., 2006]{sniffen2006guided}
Sniffen, B.~T., Harris, D.~R., and Ramsdell, J.~D. (2006).
\newblock Guided policy generation for application authors.
\newblock In {\em SELinux Symposium}.

\bibitem[Spencer et~al., 1999]{spencer1999flask}
Spencer, R., Smalley, S., Loscocco, P., Hibler, M., and Lepreau, J. (1999).
\newblock {The Flask security architecture: System support for diverse
  policies}.
\newblock In {\em USENIX Security}.

\bibitem[{Tresys}, 2014]{setools3}
{Tresys} (2014).
\newblock {SETools project page}.
\newblock \url{github.com/TresysTechnology/setools3/wiki}.
\newblock Accessed: 2015-09-29.

\bibitem[Wang et~al., 2015]{wang15easeandroid}
Wang, R., Enck, W., Reeves, D., Zhang, X., Ning, P., Xu, D., Zhou, W., and
  Azab, A. (2015).
\newblock {EASEAndroid: Automatic Policy Analysis and Refinement for Security
  Enhanced Android via Large-Scale Semi-Supervised Learning}.
\newblock In {\em USENIX Security}.

\bibitem[Zhou and Jiang, 2012]{zhou2012dissecting}
Zhou, Y. and Jiang, X. (2012).
\newblock Dissecting android malware: Characterization and evolution.
\newblock In {\em Security and Privacy}, pages 95--109. IEEE.

\end{thebibliography}

\vfill
\end{document}